\documentclass{ws-ijmpa}

\usepackage[super]{cite}
\usepackage[verbose,hypertexnames=false]{hyperref}
\usepackage{graphicx}
\usepackage{amsmath}
\usepackage{slashed}
\usepackage{amsfonts}
\usepackage{amssymb}
\usepackage{caption}
\usepackage{multirow}
\usepackage{cancel}
\usepackage{subcaption}
\usepackage{url}
\usepackage{soul}
\usepackage{calrsfs}
\usepackage{makeidx}
\usepackage{indentfirst}
\usepackage{ragged2e}
\usepackage[dvipsnames]{xcolor}
\usepackage[export]{adjustbox}
\usepackage{longtable}
\usepackage{comment}

\def\ga{\left(}
\def\dr{\right)}

\begin{document}

\markboth{T. de Oliveira, Siyuan Li, T.G. Steele}{Applying Diagrammatic Renormalization to $2^{++}$ Tensor Di-Gluonium}

%
\catchline{}{}{}{}{}
%

\title{An Application of Diagrammatic Renormalization\\ to $2^{++}$ Tensor Di-Gluonium}

\author{T. de Oliveira\footnote{tdo842@usask.ca}}

\author{Siyuan Li\footnote{siyuan.li@usask.ca}}
\author{T.G. Steele\footnote{tom.steele@usask.ca}}

\address{Department of Physics \& Engineering Physics, University of Saskatchewan \\
Saskatoon, SK, S7N~5E2, Canada}

\maketitle

\begin{history}
\received{Day Month Year}
\revised{Day Month Year}
\accepted{Day Month Year}
\published{Day Month Year}
\end{history}

\begin{abstract}
We apply the diagrammatic renormalization method to the NLO analysis of the $2^{++}$ tensor di-gluonium channel within the QCD sum-rules approach.
Diagrammatic renormalization eliminates non-local divergences directly, avoiding the construction of renormalization factors and complications arising from operator mixing in the conventional renormalization method. 
The local divergences in QCD correlation functions contribute only to subtraction terms in dispersion relations in QCD sum-rules, making it particularly well-suited for diagrammatic renormalization as the local divergences do not enter sum-rules analysis. 
We provide a detailed example of renormalizing a representative NLO diagram and perform a comprehensive comparison of all non-zero NLO diagrams for $2^{++}$ tensor di-gluonium treated with both diagrammatic and conventional operator-mixing methods. The results from both approaches are in agreement, confirming the validity of diagrammatic renormalization. 
By simplifying the renormalization process, the diagrammatic renormalization method offers a practical alternative for higher-loop analysis of gluonium states and extensions to multi-quark systems.
\end{abstract}


\ccode{PACS numbers: 03.65.$-$w, 04.62.+v}

\section{Introduction}	

As one of the hypothetical exotic hadrons, gluonia have been the subject of ongoing theoretical and experimental searches (see, e.g., Refs.~\citen{Bagan_1987,Chen_2006_tensor_lattice_predict,narison_1998,BESIII:2023wfi}). In Ref.~\citen{Li_2023_tensor}, we studied the $J^{PC} = 2^{++}$ tensor di-gluonium using the QCD Laplace sum-rules approach and improved the theoretical prediction to next-to-leading order (NLO). When evaluating the perturbative contribution to the NLO diagrams for the gluon current, we encountered challenges related to the renormalization of the Feynman diagram contributions.

In the QCD sum-rules approach, the renormalization of such systems requires the renormalization of correlation functions of composite operators (see Refs.~\citen{Shifman1979,Shifman1979a} for overview on QCD sum-rules). In the conventional renormalization method, where counter-terms are added to the Lagrangian density and additional renormalization factors are needed for composite operators, the renormalization of those composite operators results in the operator-mixing, and becomes progressively more complicated as the mass dimension of the operator increases, limiting the calculation power in QCD sum-rules applications (see Refs.~\citen{pasta,Narison_book,Collins:1984xc} for discussions on QCD renormalization). 

The diagrammatic renormalization method \cite{bogoliubov1980,Hepp:1966eg,Zimmermann:1969jj,Collins:1974da}, however, has been shown to avoid the explicit mixing of composite operators and increase the efficiency of the renormalization for scalar and vector mesonic correlation functions for light and heavy quarks, light scalar quark meson and scalar glueball mixed correlation function, for heavy-non-strange and heavy-strange diquark correlation functions, and for non-strange $0^{+-}$ tetraquark systems\cite{deOliveira:2022, deOliveira:2023hma, Ray:2022fcl}. 
In this paper, we introduce the $2^{++}$ tensor di-gluonium correlation function in Sec.~\ref{sec:tensor_correlator}. The diagrammatic renormalization method is presented and applied to the perturbative contribution of the $2^{++}$ tensor di-gluonium correlator in Sec.~\ref{sec:diagrammatic_renorm}. The correlation function renormalized via the diagrammatic renormalization method is compared to the conventional method of renormalization in Sec.~\ref{sec:conventional_renorm}.

\section{$2^{++}$ Tensor Di-gluonium Correlation Function}
\label{sec:tensor_correlator}
Using Laplace sum-rules method, Ref.~\citen{Li_2023_tensor} studies the NLO correction to $2^{++}$ Tensor Di-gluonium including both perturbative and non-perturbative contributions of the two-point correlation function~\cite{narison_1998}:
\begin{equation}\label{two-point_correlator}
    \psi_{\mu \nu \rho \sigma}^T \equiv i \int d^4 x e^{i q x}\left\langle 0\left|\mathcal{T} J_{\mu \nu}^g(x) J_{\rho \sigma}^g(0)^{\dagger}\right| 0\right\rangle =P_{\mu\nu\rho\sigma} \psi_{_T}\left(q^2\right) \,, 
\end{equation}
where $P_{\mu\nu\rho\sigma}$ is the spin-2 projection operator at $D=4+2\epsilon$ space-time dimension (see Refs.~\citen{Govaerts:1986pp}, \citen{narison_1998} and \citen{Bagan_1987}):
\begin{equation}\label{d-dimensional_projection_operator}
	P_{\mu\nu\rho\sigma}=\eta_{\mu\rho}\eta_{\nu\sigma}+\eta_{\mu\sigma}\eta_{\nu\rho}-\frac{2}{D-1}\eta_{\mu\nu}\eta_{\rho\sigma},\; \text{with}\;\eta_{\mu \nu} \equiv g_{\mu \nu}-\frac{q_\mu q_\nu}{q^2}\,,
\end{equation}
whose normalization factor is 
\begin{equation}
	 P_{\mu\nu\rho\sigma} P^{\mu\nu\rho\sigma} = 2\ga D^2-D-1 \dr\,,
\end{equation}
and $J_{\mu \nu}^g(x)$ are the gluonic currents that are built from the gluon fields~\cite{narison_1998}:
\begin{equation}\label{gluon_current}
	J_{\mu \nu}^g(x) = -G_\mu^{\alpha,a}(x) G_{\nu \alpha,a}(x)+\frac{1}{4} g_{\mu \nu} G_{\alpha \beta}^a(x) G^{\alpha \beta}_a(x)\,.
\end{equation}
The leading-order (LO) contribution to Eq.~\eqref{two-point_correlator} is a Feynman diagram with a one-loop gluonic self-energy topology. The NLO perturbative contributions to the correlation function \eqref{two-point_correlator} are listed in the second column of Tables~\ref{diagram_table_1} and \ref{diagram_table_2}. Applying QCD Laplace sum-rules methods, all local divergencies from \eqref{two-point_correlator} are removed via a Borel transform and all non-local divergences are removed via the diagrammatic renormalization method.

\section{Diagrammatic Renormalization}
\label{sec:diagrammatic_renorm}
In the diagrammatic renormalization method (see, e.g., Refs.~\citen{Collins:1984xc, deOliveira:2022} for a review and applications of the method) each Feynman diagram $G$ from the perturbative expansion of Eq.~\eqref{two-point_correlator} is individually renormalized by subtracting counter-term diagrams $C(G)$ from the bare diagram $U(G)$. To construct each $C(G)$, isolate a sub-diagram from $G$ (i.e., a subset of lines that contains at least one loop), extract the divergent term $\gamma$ from the sub-diagram (the sub-divergence), and then replace the sub-diagram by $\gamma$ in $G$. The renormalized diagram $R(G)$ is obtained by subtracting the counter-term diagrams associated with each sub-diagram, $C_{\gamma}(G)$, from the bare diagram $U(G)$, 
\begin{equation}
R(G)=U(G)-\sum_{\gamma}C_{\gamma}(G)\,.
\label{diag_ren_formula}
\end{equation}
After each diagram $G$ from the correlation function is renormalized through Eq.~\eqref{diag_ren_formula}, the sum of all $R(G)$ is the renormalized correlation function with all non-local divergences removed from Eq.~\eqref{two-point_correlator} (local divergencies are removed in QCD sum-rule methods) and with coupling and mass parameters interpreted as their renormalized versions. An example of the $2^{++}$ tensor di-gluonium NLO perturbative evaluation of the bare diagram for $\psi_{_T} (q^2)$ along with the diagrammatic renormalization procedure is provided in Sec.~\ref{sec:example_diagram_b}.

\subsection{Example: Gluon Self-Energy with Gluonic Loop}
\label{sec:example_diagram_b}

\begin{figure}[h]
\centerline{\includegraphics{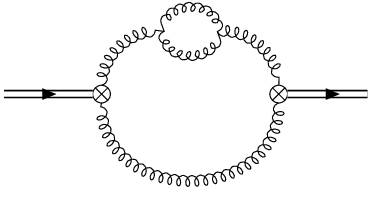}}
\caption{NLO gluon-self energy with gluonic loop non-zero contribution to the correlation function \eqref{two-point_correlator}, where the curly lines stands for gluon fields, the $\otimes$ denotes the composite operator insertion, and the double line represents the external momentum $q$.}
\label{figure_1}
\end{figure}

The non-zero contribution to Eq.~\eqref{two-point_correlator} for the gluon self-energy topology (diagram ii) in Table~\ref{diagram_table_1}) is shown in Fig.~\ref{figure_1} and is given by (the applied Feynman rules can be referred to in Refs.~\citen{pasta,cheng_li,li_2025_thesis}): 
\begin{equation}\label{c_bare}
                  \Pi^{(\mathrm{ii})}_{\rm bare}(Q^2 \equiv -q^2)
            = \frac{19 \, \alpha_s\, Q^4}{1600 \pi^3} L \left( 10 L -\frac{2642}{57} + \frac{10}{\epsilon} \right)\,, \, \, 
              L =  \log\left(\frac{Q^2}{\nu^2}\right)\,, 
\end{equation}
where the result is in Feynman gauge, $\nu$ is the $\rm{MS}$ scheme renormalization scale, and local divergences (polynomials in $Q^2$) are omitted from Eq.~\eqref{c_bare}.

\begin{figure}[h]
\centerline{\includegraphics{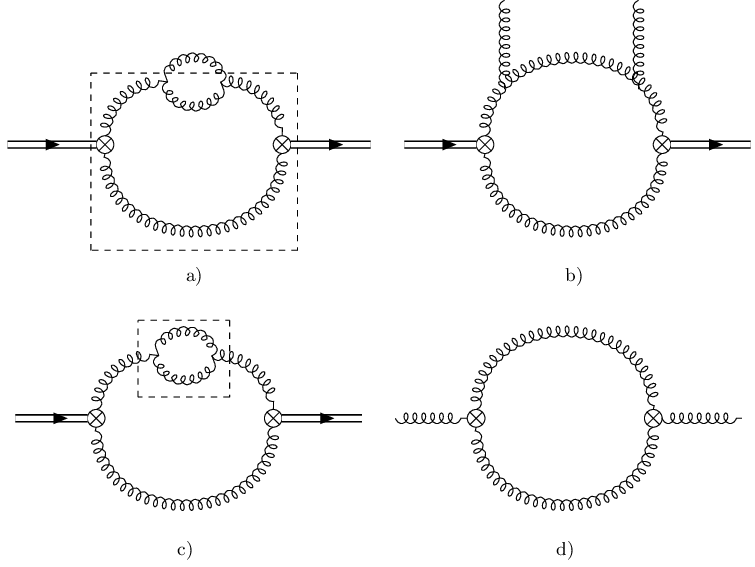}}
\caption{Sub-diagrams $b)$ and $d)$ extracted from Fig.~\ref{figure_1}. Isolating the loop indicated by the dashed boxes in diagrams $a)$ and $c)$ originates sub-diagrams $b)$ and $d)$, respectively. The curly lines represent the gluon fields, $\otimes$ denotes the composite operator insertion, and the double line represents the external momentum $q$.}
\label{figure_2}
\end{figure}

\begin{figure}[h]
\centerline{\includegraphics{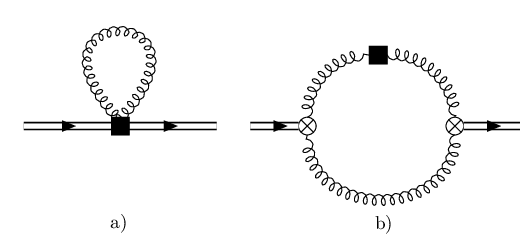}}
\caption{Counter-term diagrams generated by the sub-diagrams of Fig.~\ref{figure_2} $b)$ and $d)$. The curly lines represent the gluon fields, $\otimes$ denotes the composite operator insertion, the square $\blacksquare$ represents the sub-divergence insertion, and the double line represents the external momentum $q$.}
\label{figure_3}
\end{figure}

The diagram in Fig.~\ref{figure_1} has two sub-diagrams, represented in the third column of Table \ref{diagram_table_1} and in Figs.~\ref{figure_2} $b)$ and $d)$. Figs.~\ref{figure_2} $b)$ and $d)$ are the loops indicated by the dashed boxes in Figs.~\ref{figure_2} $a)$ and $c)$, respectively. Once the sub-diagrams are isolated, the counter-term diagrams (see the fourth column in Table \ref{diagram_table_1}) are generated by replacing the dashed boxes in Figs.~\ref{figure_2} $a)$ and $c)$ by the corresponding divergent term from Figs.~\ref{figure_2} $b)$ and $d)$, resulting in Figs.~\ref{figure_3} $a)$ and $b)$, respectively.

The counter-term in Fig.~\ref{figure_3} $a)$ is a massless tadpole.
The loop integral of such topology is zero in dimensional regularization~\cite{pasta}.
Even in the case of a massive tadpole, the external momentum $q$ does not enter the loop integral for Fig.~\ref{figure_3} $a)$. Hence, the tadpole diagram corresponds to a dispersion-relation subtraction constant in Eq.~\eqref{two-point_correlator} and does not have a physical contribution to QCD sum-rules, representing a local divergence. This type of counter-term diagram also shows up for the bare diagrams iii), iv), v) and vi) in Tables~\ref{diagram_table_1} and \ref{diagram_table_2}.

For the counter-term diagram in Fig.~\ref{figure_3} $a)$, the sub-divergence $\gamma$ insertion, represented by the square $\blacksquare$, from sub-diagram $d)$ in Fig.~\ref{figure_2} is given by
\begin{equation}\label{c_SE_g_subdiv}
    \gamma = i \delta^{ab} \frac{g^2}{64 \pi^2 \epsilon} \left( 22q^\mu p^\nu - 19p^2g^{\mu\nu}  \right)\,,
\end{equation}
where the Latin letters represent color indices. Note that the $\frac{1}{\epsilon}$ divergence subtraction are performed in MS scheme. So the renormalized result will need conversion to $\overline{\mathrm{MS}}$ scheme using $\nu^2 \rightarrow \exp(\gamma_{_E})\nu^2/ 4\pi$. 
The resulting contribution $\Pi_{\rm ct}^{(\mathrm{ii})}$ from Fig.~\ref{figure_3} $b)$ is then
\begin{equation}\label{c_sub}
    \Pi^{(\mathrm{ii})}_{\rm ct}( Q^2 \equiv -q^2) 
     = \frac{19 \, \alpha_s\, Q^4}{1600 \pi^3} L \left( 5 L -9 +\frac{10}{\epsilon} \right), \; L =  \log\left(\frac{Q^2}{\nu^2}\right).
\end{equation}
Note that the non-local divergent $L/ \epsilon$ terms from Eq.~\eqref{c_bare} and \eqref{c_sub} are identical.
According to the diagrammatic renormalization method, the renormalized diagram can be found by subtracting the counter-term diagram contribution given in Eq.~\eqref{c_sub} from the bare diagram contribution given in Eq.~\eqref{c_bare}. Therefore, for the diagram in Fig.~\ref{figure_1}, the renormalized result is
\begin{equation}\label{c_renorm}
\begin{aligned}
    \Pi^{(\mathrm{ii})}_{\rm renorm}(Q^2)&=\Pi^{(\mathrm{ii})}_{\rm bare}(Q^2)-\Pi^{(\mathrm{ii})}_{\rm ct}(Q^2) \\
    &= \frac{\alpha_s Q^4}{4800\pi^3} L
    \left( 285 L - 2129 \right),\; \; \;L =  \log\left(\frac{Q^2}{\nu^2}\right) \,.    
\end{aligned}
\end{equation}

\subsection{Renormalized Correlation Function}
To obtain the complete NLO $2^{++}$ tensor di-gluonium renormalized correlation function, one must sum the contributions from all the individual NLO renormalized diagrams shown in Tables \ref{diagram_table_1} and \ref{diagram_table_2}:
\begin{equation}\label{nlo_renorm}
    \begin{aligned}
		\Pi^{NLO}_{\rm renorm}(Q^2)  &= \sum_{m=a}^{g} \Pi^{(m)}_{\rm renorm}
		\\&= \frac{\alpha_s}{1800\pi^3} Q^{4} L
		\left[ \left(101 - 15 L \right) n_f +150  \right] ,
		\;L =  \log\left(\frac{Q^2}{\nu^2}\right)\,
	\end{aligned}
\end{equation}
for $n_f$ flavors.
\begin{table}[ht]
\tbl{NLO (non-zero) bare diagrams, sub-diagrams, and counter-term diagrams for the $2^{++}$ tensor di-gluonium in the diagrammatic renormalization procedure. Curly lines represent the gluon fields, dotted lines represent the ghost field, solid lines represent the quark field, $\otimes$ is the composite operator insertion, $\blacksquare$ is the sub-divergence insertion, and the double line represents the external momentum $q$. 
Diagram symmetry-related cases are taken into account and incorporated into the final results.
\label{diagram_table_1}}
{\includegraphics[height=1.\linewidth,angle=90]{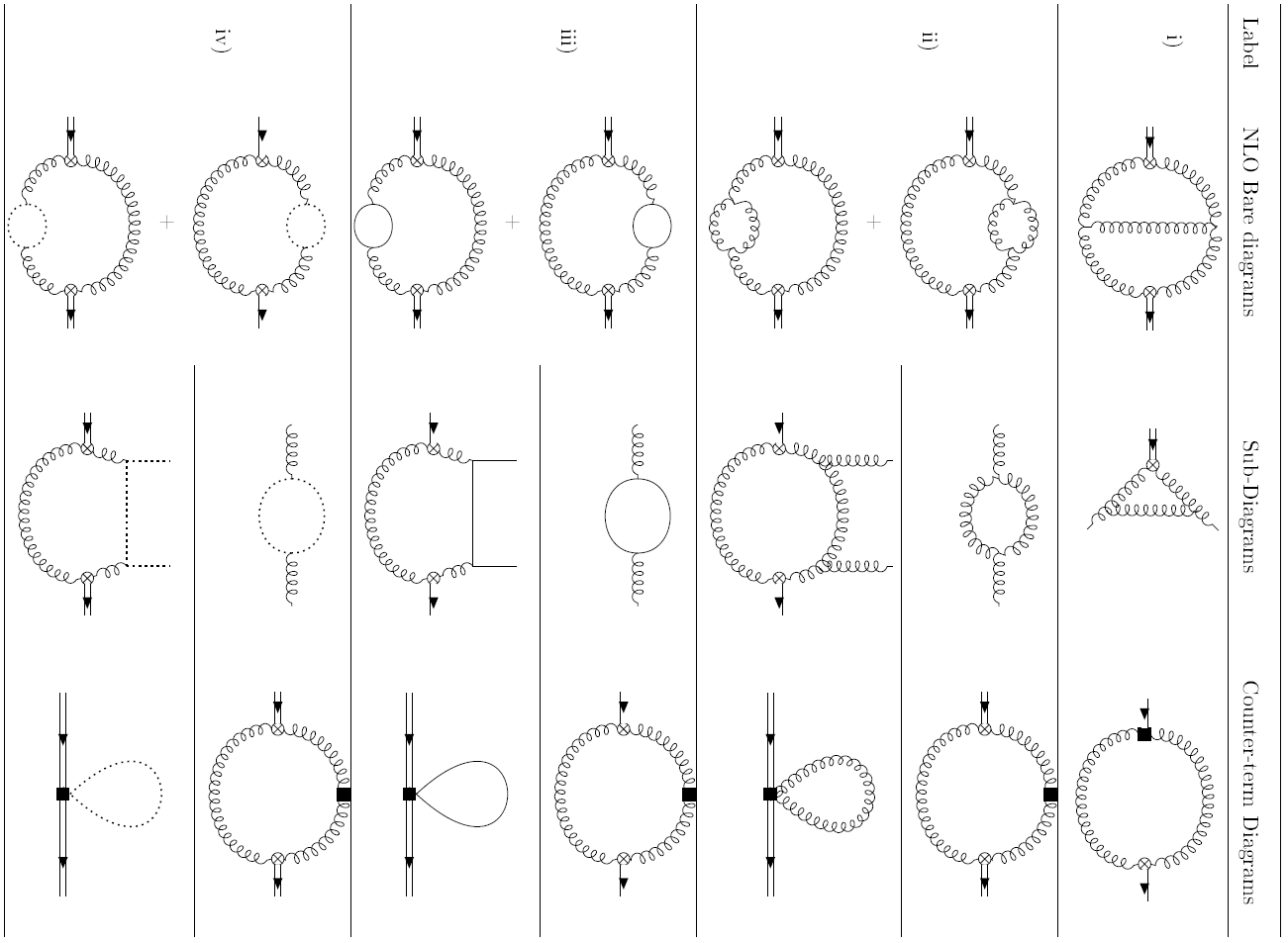}}
\end{table}

\begin{table}[ht]
\tbl{(Continued)
NLO (non-zero) bare diagrams, sub-diagrams, and counter-term diagrams for the $2^{++}$ tensor di-gluonium in the diagrammatic renormalization procedure. Curly lines represent the gluon fields, dotted lines represent the ghost field, solid lines represent the quark field, $\otimes$ is the composite operator insertion, $\blacksquare$ is the sub-divergence insertion, and the double line represents the external momentum $q$. 
Diagram symmetry-related cases are taken into account and incorporated into the final results.
\label{diagram_table_2}}
{\includegraphics[height=1\linewidth,angle=90]{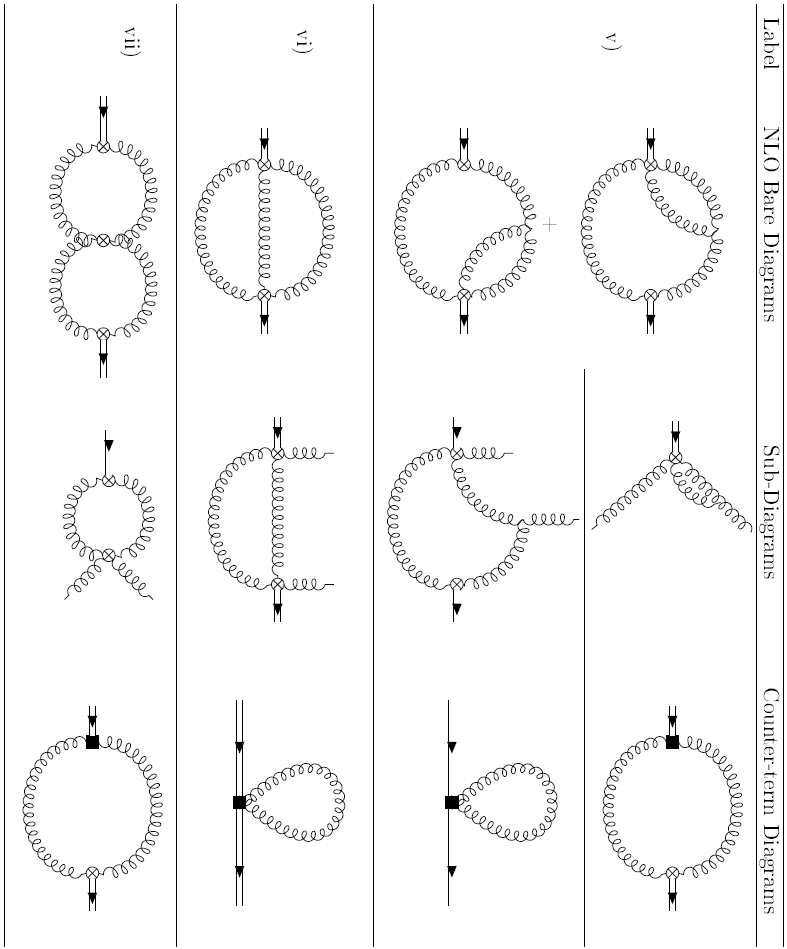}}
\end{table}

The contributions of each diagram --- bare, counter-term, and renormalized --- are summarized in Table~\ref{result_table}, while the corresponding diagram representations are presented in Tables~\ref{diagram_table_1} and \ref{diagram_table_2}. 
Diagram symmetry-related cases are taken into account and incorporated into the final results in Table \ref{result_table}.

\begin{table}[h]
\caption{NLO (perturbative) bare, subtraction, and counterterm diagrams for the $2^{++}$ tensor di-gluonium contribution to $\psi_{_T} (q^2)$ in Eq.~\eqref{two-point_correlator}, evaluated using both the conventional and diagrammatic renormalization methods.
$A$ and $B$ are defined by $\psi_{_T} (Q^2) = \left(\frac{\alpha_s}{\pi}\right)^3 (\frac{Q^4}{16}) L \left( A L +B\right)$ for each diagram's contribution. 
$n_f$ stands for number of flavors. 
Diagram viii) is used exclusively in the conventional renormalization approach, as detailed in Sec.~\ref{sec:conventional_renorm}. All symmetry-related diagram cases have been evaluated and incorporated into the final results.
\label{result_table}}
\renewcommand{\arraystretch}{1.7}
\begin{tabular}{c c c c c c c}
	\hline
	Label  & \multicolumn{2}{c}{Bare} & \multicolumn{2}{c}{Counter-term} & \multicolumn{2}{c}{Renormalized}\\
	\cline{2-7}
	& A & B &  A & B & A & B \\
	\hline
	i) & $-\frac{7}{30}$ & $\frac{3107}{1800}-\frac{7}{30\epsilon}$ & $-\frac{7}{60}$ & $\frac{521}{225}-\frac{7}{30\epsilon}$ & $-\frac{7}{60}$ & $-\frac{1151}{1800}$\\
	ii) & $\frac{19}{10}$ & $-\frac{1321}{150}+\frac{19}{10\epsilon}$  & $\frac{19}{20}$  & $-\frac{171}{100}+\frac{19}{10\epsilon}$ & $\frac{19}{20}$ & $-\frac{2129}{300}$\\
	iii) & $-\frac{4}{15}n_f$ & $\frac{256}{225}n_f-\frac{4}{15\epsilon}n_f$  & $-\frac{2}{15}n_f$  & $\frac{6}{25}n_f-\frac{4}{15\epsilon}n_f$ & $-\frac{2}{15}n_f$ & $\frac{202}{225}n_f$\\
	iv) & $\frac{1}{10}$ & $-\frac{79}{150}+\frac{1}{10\epsilon}$  & $\frac{1}{20}$  & $-\frac{9}{100}+\frac{1}{10\epsilon}$ & $\frac{1}{20}$ & $-\frac{131}{300}$\\
	v) & 0 & -1 & 0 & 0 & 0 & -1\\
	vi) & $-\frac{31}{10}$ & $\frac{2887}{200} - \frac{31}{10\epsilon}$ & $-\frac{31}{20}$ & $\frac{128}{75}-\frac{31}{10\epsilon}$& $-\frac{31}{20}$ & $\frac{7637}{600}$ \\
	vii) & $\frac{4}{3}$ & $-\frac{40}{9}+\frac{4}{3\epsilon}$ & $\frac{2}{3}$ & $-\frac{20}{9} + \frac{4}{3\epsilon}$ &$\frac{2}{3}$ & $-\frac{20}{9}$\\
        viii) & $\frac{2}{15}n_f$ & $-\frac{6}{25}n_f + \frac{4}{15\epsilon}n_f$ & N/A & N/A & N/A & N/A\\
    \hline
    Total & $-\frac{2}{15}n_f$ & $\frac{202}{225}n_f +\frac{4}{3}$ & & & $-\frac{2}{15}n_f$ & $\frac{202}{225}n_f +\frac{4}{3}$\\
    \hline
\end{tabular}
\end{table}

\section{Comparison to Conventional Renormalization Method}\label{sec:conventional_renorm}
The diagrammatic renormalization method is particularly convenient for handling the renormalization of individual diagrams. However, when determining the overall NLO contribution to the $2^{++}$ tensor di-gluonium, the conventional approach involves incorporating renormalization-induced diagrams.

As discussed in Ref.~\citen{Bagan_1987}, the renormalized meson-gluonium currents take the following operator-mixing forms:
\begin{equation}
	\begin{aligned}
		&J^{\mu\nu}_{q,R} = Z_{11} J^{\mu\nu}_{q,B} + Z_{12} J^{\mu\nu}_{g,B}\,,\\
		&J^{\mu\nu}_{g,R} = Z_{21} J^{\mu\nu}_{q,B} + Z_{22} J^{\mu\nu}_{g,B}\,,
	\end{aligned}
\end{equation}
where the renormalization constants $Z_{ij}$ are in $\overline{\text{MS}}$ scheme, and $J^{\mu\nu}_{q,B}$ is the bare quark current. To $\mathcal{O}(\alpha_s/\pi)$, the only renormalization constant needed is $Z_{22}$ that connects the bare gluon current $J^{\mu\nu}_{g,B}$ to renormalized gluon current $J^{\mu\nu}_{g,R}$:
\begin{equation}\label{Z22}
	Z_{22} = 1 + R_{22}, \text{ with } R_{22}= -\frac{\alpha_s}{6 \pi \epsilon} n_f\,.
\end{equation}
$R_{22}$ will construct a new diagram term, called the renormalization-induced diagram, to renormalize the divergences from the carried bare diagrams. $R_{22}$ will be embedded in a term which is proportional to the current-gluon vertex, represented as the shaded circle in Fig.~\ref{figure_4}. Due to the symmetry of the tensor gluonium LO diagram, there are two renormalization-induced diagrams, represented in Figs.~\ref{figure_4} $a)$ and $b)$, both yielding identical contributions. The sum of both contributions gives:
\begin{figure}[h]
\centerline{\includegraphics{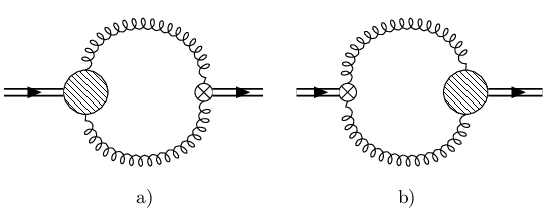}}
\caption{Renormalization induced-diagrams in the conventional operator-mixing approach. Curly lines represent the gluon fields, $\otimes$ denotes the composite operator insertion, the shaded circle represents $R_{22}$ contribution, and the double line represents the external momentum $q$.}
\label{figure_4}
\end{figure}

\begin{equation}\label{diagram_h_renorm_induce}
	\begin{aligned}
		\Pi^{(viii)}_{\rm bare} (Q^2 \equiv -q^2) = \frac{\alpha_s Q^4 }{600 \pi^3} n_f L \left( 5L +\frac{10}{\epsilon} - 9 \right) ,
		\;L =  \log\left(\frac{Q^2}{\nu^2}\right)\,.
	\end{aligned}
\end{equation}
The renormalized $2^{++}$ tensor di-gluonium is obtained by summing the NLO bare diagrams and the renormalization-induced diagrams determined by $R_{22}$. The result from the conventional renormalization method matches the one derived using the diagrammatic renormalization method as given in Eq.~\eqref{nlo_renorm}, and shown in the last row of Table~\ref{result_table}.
This agreement serves as a strong validation of the diagrammatic renormalization method when applied to the $2^{++}$ tensor di-gluonium.

\section{Conclusion}
Diagrammatic renormalization offers an alternative to conventional renormalization techniques for QCD correlation functions. This approach effectively eliminates non-local divergences, thereby circumventing the intricate construction of renormalization factors for composite operators—particularly those involving operator mixing. It is especially well-suited for QCD sum-rule analyses, where local divergences manifest solely as subtraction terms in the dispersion relation and thus do not contribute to the sum-rule itself. As a result, the renormalization process is significantly simplified, requiring attention only to the logarithmic (non-local) divergences.

In this paper, we provided a detailed example within the QCD sum-rule framework by applying diagrammatic renormalization to a NLO diagram of the $2^{++}$ tensor di-gluonium in Sec.~\ref{sec:diagrammatic_renorm}. A comprehensive comparison of all relevant NLO diagrams renormalized via both diagrammatic and conventional methods is presented in Sec.~\ref{sec:conventional_renorm} and summarized in Table~\ref{result_table} (see also Refs.~\citen{Li_2023_tensor,li_2025_thesis}). We find that diagrammatic renormalization yields results in full agreement with the conventional approach. By bypassing the complexities of operator mixing, diagrammatic renormalization emerges as a viable and efficient method for studying higher-loop corrections in gluonium and, more broadly, multi-quark systems within QCD sum-rules\cite{deOliveira:2022}.

\section*{Acknowledgments}
TGS is grateful for research funding from the Natural Sciences and Engineering Research Council of Canada (NSERC).

\section*{ORCID}

\noindent T. de Oliveira- \url{https://orcid.org/0000-0003-3976-1272}

\noindent Siyuan Li - \url{https://orcid.org/0009-0002-9582-4864}

\noindent T. G. Steele - \url{https://orcid.org/0000-0003-1716-0783}

\bibliography{reference}
\bibliographystyle{ws-ijmpa}

\end{document}